# Opening Pandora's box: Paper mills in conference proceedings

**Anna Abalkina, research fellow,** Free University of Berlin, anna.abalkina@fu-berlin.de (https://orcid.org/0000-0003-1469-4907)

**Marie Kunešová, researcher,** NTIS Research Centre, University of West Bohemia in Pilsen, Czechia, mkunes@ntis.zcu.cz (https://orcid.org/0000-0002-7187-8481)

**Yagmur Ozturk, PhD student,** Univ. Grenoble Alpes, CNRS, Grenoble INP, LIG, 38000 Grenoble, France, yagmur.ozturk@univ-grenoble-alpes.fr (https://orcid.org/0000-0002-2843-8990)

**Solal Pirelli, research engineer,** EPFL, solal.pirelli@epfl.ch (https://orcid.org/0009-0003-4336-1316)

## Abstract

Paper mills are a growing threat to the integrity of science, yet their penetration in conference proceedings remains underexplored despite conferences being more important than journals in some scientific subfields. This study aims to identify papers in conference proceedings whose titles have been offered for sale on social media platforms. We collected publication offers for more than 4,000 unique paper titles or topics from more than 200 social media channels and used semi-automated methods along with human assessment to match offers with papers published in IEEE conference proceedings. We identified 2,062 papers in 366 IEEE conference proceedings, accounting for up to 23.51% of an individual conference. These problematic papers are co-authored by more than 7,500 researchers from over 4,000 affiliations in 56 countries. The identified papers demonstrate collaboration anomalies, high diversity of affiliations per paper, citation manipulation, a predominance of six-author papers, and content-based irregularities. Our findings show that paper mills are a large, organized, and often public market that commercializes scientific misconduct, not limited to papers, but infiltrating multiple parts of the research ecosystem.

## Introduction

Maintaining research integrity is crucial for the credibility of science. A key threat to research integrity is the rise of paper mills, commercial entities that sell co-authorship slots and/or violate the peer-review process (Abalkina et al., 2025; Hvistendahl, 2013; Parker et al., 2024; Richardson et al., 2025a; Ro & Leeming, 2025; Van Noorden, 2023). There is evidence that papers originating from paper mills are associated with scientific misconduct (Abalkina, 2023; McCook, 2016), violation of the peer review process (Abalkina & Bishop, 2023; Day, 2022; Kersjes, 2025; Matusz et al., 2025), citation manipulation (Candal-Pedreira et al., 2024), tortured phrases (Cabanac et al., 2021), image duplication (Aquarius et al., 2025a; Bik et al., 2016; Christopher, 2021), poor scientific analysis (Suchak et al., 2025), and various errors (Hackenberg et al., 2025; Park et al., 2022; Richardson et al., 2025b). Paper mills have become a growing concern in discussions about trust in science and the reliability of scientific results.

Paper mills reach their clients through various channels, including unsolicited emails, websites, promotional seminars at universities (Abakina, 2023), social media, and messaging apps. Paper mills have been reported to advertise their services on social media such as WhatsApp and Telegram (Basava, 2025; Hidalgo, 2023; Richardson et al., 2026). Such services include “authorship for sale”, i.e., adding someone as an author of a paper they did not write in exchange for money. This violates the authorship policies of virtually every scientific publisher and institution.

The majority of studies exploring paper mills focus on journal papers, while conference proceedings have received less attention. This oversight is significant because in some fields, particularly computer science, conference proceedings serve as the primary venue for disseminating original research results (Freyne et al., 2010; Kochetkov et al., 2022). Furthermore, conferences are central to academic communication, as they allow researchers to present their findings, discuss them, receive feedback, and develop professional networks.

There is growing evidence of paper mills targeting conference proceedings, which can be vulnerable to the publication of problematic papers for several reasons. First, there is concern about low-quality peer review in some conference proceedings, lack of transparency in the publication process, and low barriers to publishing papers (Joelving, 2023; Kulczycki et al., 2024; Lang et al., 2019). Second, proceedings are often indexed in international databases and used in institutional or national research evaluation systems, making them attractive targets for manipulation. At the same time, with the increasing number of publications, there is pressure on publishers to maintain quality control and scrutiny in their journals. Such standards are unfortunately not always applied to published conference proceedings, especially when the conference was organized by a third party, a scenario similar to guest-edited journal issues. Thus, while there are highly reputable conferences with strict acceptance criteria and strong competition, others allow fast publication with unreliable quality control.

These concerns are particularly relevant to the Institute of Electrical and Electronics Engineers (IEEE), which, despite its name, also publishes venues in other fields such as computer science. Many IEEE-published conferences are among the most important venues in their fields, such as ICSE in software engineering or ICASSP in signal processing. Besides its own conferences, IEEE also financially or technically sponsors and co-sponsors conferences organized by others, including local IEEE sections and subsections.

IEEE is one of the leading publishers in terms of retractions (Lendvai & Sasvári, 2025) and is associated with a number of bulk retractions. For instance, IEEE has retracted papers for incoherent text (IEEE, 2025) and violation of its authorship policy (IEEE, 2024). Swart et al. (2025) analyzed conference proceedings to identify concentrations of articles containing tortured phrases, distorted scientific terminology that often signals the use of a paraphrasing tool to evade plagiarism detection. Their distributional analysis showed that IEEE accounts for 88% of the articles in their dataset containing such phrases, suggesting possible coordinated production mechanisms consistent with paper mill activity.

This study investigates paper mills in conference proceedings, with a particular focus on IEEE due to its dominance in retraction rankings. Per Lendvai & Sasvári (2025), IEEE represents 34% of retractions indexed in Scopus between 2001 and 2024, while according to the Retraction Watch database (accessed June 29, 2026), IEEE accounts for 15.5% of all retractions in the database, making it the second largest in this ranking.

By identifying papers offered for sale online and tracing their subsequent publication, this study contributes to three areas of research.

First, we expand empirical knowledge about paper mills in conference proceedings. Second, we advance the understanding of authorship-for-sale mechanisms by analyzing co-authorship slot offers, which have so far been examined only in a limited number of individual cases. Third, we provide the first systematic exploration of paper mill ecosystems across social media platforms.

We use the metaphor of Pandora's box to examine paper mills in conferences, a problem whose scale is visible only when the box is opened. Once exposing authorship-for-sale operations and fraud, they can no longer be ignored.

# Overview of Paper Mills

A conservative estimate suggests that at least 400,000 papers in journals and conferences potentially originate from paper mills, reaching approximately 3% in the biomedical literature (Van Noorden, 2023), and the share of such problematic papers has been growing over the last decade. In certain scientific disciplines, the prevalence is much higher. For example, cancer research literature includes up to 14% of potentially problematic publications in 2024, with even higher shares in gastric cancer (22%) and bone cancer (20%) (Scancar et al., 2026). Similarly, up to 40% of animal studies on hemorrhagic stroke were identified as problematic (Aquarius et al., 2025b). Fields characterized by high publication volume, standardized experimental designs, and intense funding competition may be particularly attractive to paper mills, as manuscripts can be templated and produced at scale.

The expansion of paper mills is closely linked to the structural pressures in the academic system shaped by a "paper culture" (Alhuay-Quispe & Yance-Yupari, 2025) and the "publish or perish" logic in modern research (Goddiksen et al., 2023). Publication-based research evaluation coupled with monetary rewards for publication (Quan et al., 2017) in certain indexes, such as Scopus or Web of Science, has created both requirements and incentives to increase the publication level. In such environments, publications function not only as scientific contributions but also as measurable indicators of career success and funding eligibility. These conditions have enabled new forms of contract cheating facilitated by paper mills (Byrne & Christopher, 2020).

For the buyers of such services, these structural incentives can be interpreted through established theoretical frameworks. From a rational choice perspective (Becker, 1968), engagement with paper mills can be viewed as a cost-benefit calculation in which researchers weigh the potential rewards of publication, such as financial bonuses, career advancement, or institutional recognition, against the perceived severity of sanctions. When detection mechanisms are weak or enforcement is inconsistent, the expected costs of misconduct appear relatively low. Strain theory (Merton, 1938) provides a complementary interpretation. Academic systems strongly value publication in indexed outlets as a marker of success, yet access to the resources necessary to achieve this goal, such as funding and infrastructure, is unevenly distributed. When institutional expectations exceed available means, some researchers may adopt what Merton described as "innovation," pursuing culturally valued goals through illegitimate mechanisms, including the purchase of authorship through paper mills.

At the same time, publishing is also growing rapidly, with a disproportional increase in certain publication types and journals, which can be more vulnerable to manipulation, including special issues, predatory publishing, and conference proceedings (Hanson et al., 2024). Compared to 2015, the number of journal articles indexed in Scopus in 2024 grew by 58%, conference papers by 25%, and retracted articles increased 2.5-fold. Article processing charges (APCs) give publishers a strong incentive to expand their portfolio and increase publication output. This growth puts a strain on publishing, resulting in difficulties in finding reviewers, pressure to shorten review times, and, thus, threatening quality control. These conditions facilitate the penetration of paper mills.

Empirical research on paper mills has developed along several directions. First, studies have examined papers within certain journals. These studies are co-authored mainly by the journal's editors or publisher (Fintoft et al., 2023; Hackett & Kelly, 2020; Matusz et al., 2025) or include a systematic analysis of problematic cases (Aquarius et al., 2025a). Second, researchers analyzed paper mill operations (also mentioned as questionable ghostwriting practices and authorship-for-sale services) in national contexts, including China (Hvistendahl, 2013), Peru (Alhuay-Quispe & Yance-Yupari, 2025), Kenya (Kingori, 2021), and Russia (Abalkina, 2023). Third, certain research fields appear particularly targeted, especially biomedical disciplines and psychology (Byrne & Christopher, 2020; Abalkina & Bishop, 2023). Fourth, a large and growing body of literature focuses on developing detection methods, including image forensics, network and scientometric approaches, statistical similarity and clustering techniques, automated text analysis, and identification of field-specific issues, such as incorrect nucleotide sequences (Labbé et al., 2019; Porter & McIntosh, 2024; González-Márquez et al., 2024; Scancar et al., 2026). Finally, another approach involves the detection of online authorship-for-sale titles, which provides direct evidence of a connection between a published paper and paper-mill activities (Abalkina, 2023). However, identification of paper-mill papers remains underexplored, even though many offers are publicly published online on different websites or in social media. This is particularly true of conference papers.

# Methodology and Data

## Channels

Authorship-for-sale services advertise on social media, in instant messaging services, and on online websites. Many of these channels are public, meaning anyone can find them by name to view their contents or access them through invitation links publicly posted in other channels. Contrary to what one might expect, the nature of these services and their advertisements is rarely hidden. Channel names can be blunt, such as “IEEE Conference / Scopus / Web of Science publications” or “Publication & co-authors”, or euphemistic, such as “Research Assistance” or “PhD admission/PhD Assistance and Guidance”. These channels regularly post offers for sale that are even more direct, using advertisements such as “call for author positions”, “conference accepted papers available”, or “add your name in research paper”, listing titles or keywords alongside the positions for sale and their prices. Previous evidence has demonstrated that advertised titles may correspond exactly or almost exactly to the title of the final published article (Abalkina, 2023; Naddaf, 2026). These services also often publish confirmation of paper acceptance and may offer to handle conference presentations (Supplementary material, Appendix 1). Scopus indexing is explicitly advertised, typically as “100%” or “guaranteed”. Figure 1 contains a representative advertisement.

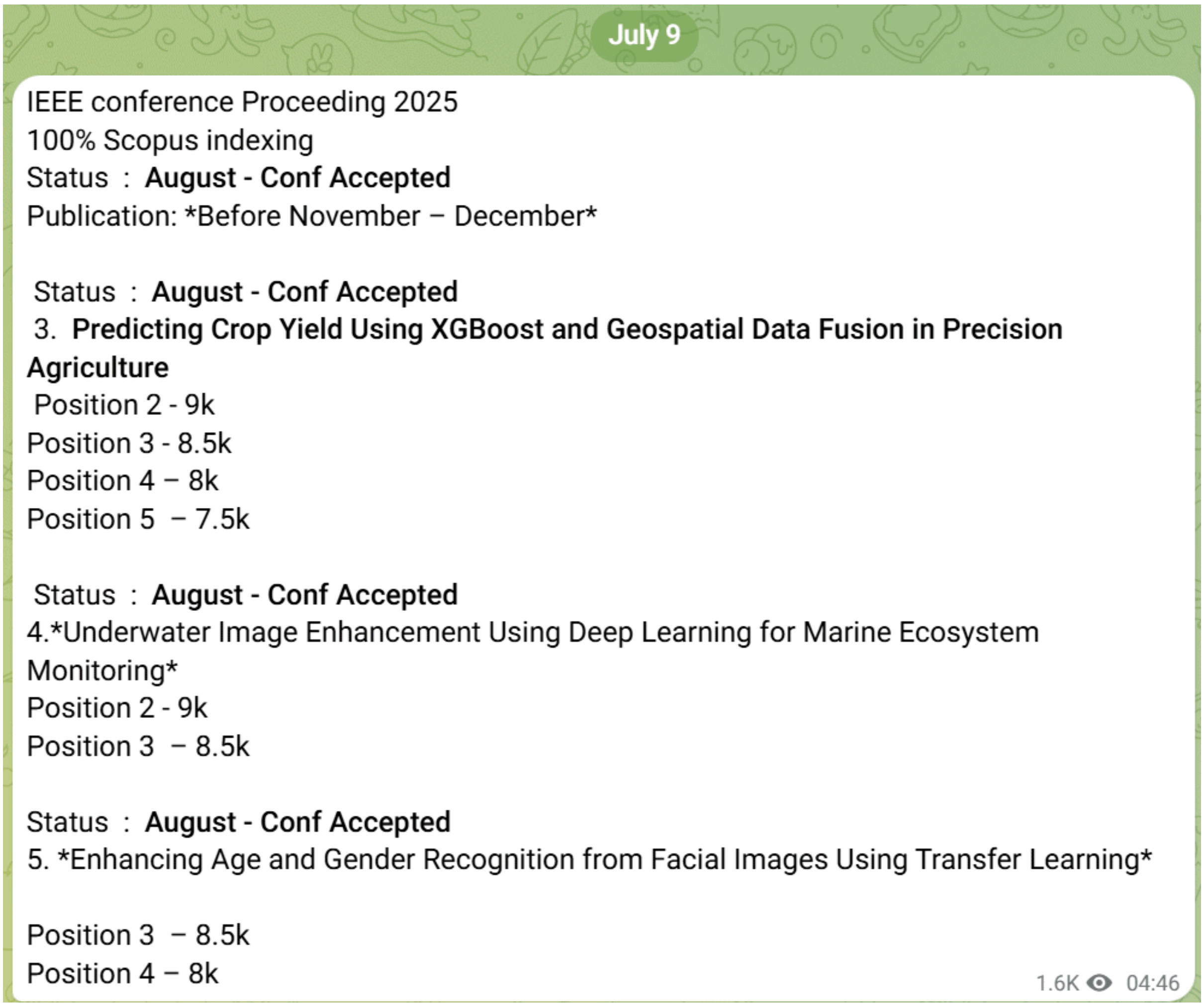

Figure 1. A representative advertisement on Telegram. Prices are in INR in this channel.

We searched for channels selling conference paper authorship on social media, on instant messaging services, and through search engines. We found over 200 channels (re-)posting such offers, including specialized authorship-for-sale accounts with dozens to thousands of followers, as well as large publication support groups with up to 50,000 members. The cumulative audience of the channels we identified exceeds 600,000 people.

In addition to conference and journal papers, authorship-for-sale accounts also offer author positions on patents, books, and book chapters, as well as a range of other services. We spotted advertisements for citation boosting, thesis and student project writing, plagiarism and AI detection reports, “plagiarism removal”, programming project implementation, journal editorial board memberships, book editor positions, and award certificates. Occasionally, these services are linked to immigration opportunities, such as assistance with meeting the requirements to get a US visa or securing nursing jobs in Germany, including visa applications, German-language training, as well as travel and settlement guidance (Supplementary material, Appendix 2).

Larger channels tend to be more generic in nature and include posts by various authorship-for-sale services alongside other related subjects, such as individuals asking for a copy of a paywalled scientific article. In one particularly stunning example, we found authorship for IEEE conference papers being openly sold in *the official Facebook group of an IEEE region* (Supplementary material, Appendix 3).

Generic channels also occasionally include authorship-for-sale offers posted by individuals who appear to be unaffiliated with any service, framed as seeking a sponsor who would pay the article’s publishing fees in exchange for an author position.

## Collecting offers

Starting in July 2025, we searched for authorship-for-sale offers for IEEE conference papers on Facebook, X (formerly Twitter), Bluesky, Instagram, Telegram, WhatsApp, LinkedIn, YouTube, and online websites. We collected 5,684 offers for IEEE publications, reduced to 4,407 unique titles or sets of keywords after deduplication of identical or near-identical titles (e.g., with alternate spellings). These offers were advertised between February 23, 2021, and December 31, 2025, and were posted both individually and in blocks of up to 274 offered titles within a single post.

We only observed the messages posted in the identified channels and did not solicit any offers or directly interact with any humans during this search. However, we did get added to some private authorship-for-sale groups on WhatsApp without asking, probably because we had previously joined similar public ones.

Almost all the collected offers are in English, with a few in other languages that we machine-translated. We stored all offers in a spreadsheet that includes the offered title or keywords, the social media channel, the posting date, the number of authorship slots offered, and a direct link to the offer when available. We also collected screenshots of offers.

While this study focuses on offers related to IEEE conferences, we also found offers for titles in IEEE journals and conference proceedings from other publishers, which we stored in a separate spreadsheet. We did not systematically search for other publishers, which is why these offers are a secondary output of our method, with only 886 offers compared to the 4,407 ones for IEEE.

## Overview of the collected offers

The typical offer for authorship of conference papers includes the publisher, a guarantee of Scopus indexing, and tentative titles or, more rarely, sets of keywords. Available authorship slots for sale are then listed, often with a price per slot. Publication times, when mentioned, range from 1 to 6 months. Some advertisements include additional information such as the precise or rough date of expected publication. In some cases, the papers whose authorship is being sold have already been accepted, as in Figure 1, and, in one case we found, already presented at a conference (Supplementary material, Appendix 4).

The number of co-authorship slots is usually six (Supplementary material, Appendix 5). We do not have a concrete explanation for this phenomenon, although we suspect that it may be related to how authors are displayed on the IEEE Xplore platform. At the time of writing, the author list at the top of each paper's webpage only shows the first six authors, while the full authorship list is located below the paper. This decreased prominence of authors in positions 7+ could lead to a lower interest in such author slots. Additionally, a few conferences explicitly limit the number of authors per paper to six.

Price mainly depends on the position and the country of the seller, with the highest price for the first-author position and each subsequent position being cheaper. The price range for conference authorships, listed in national currencies or US dollars, ranges from $43 to $400 for the first authorship, and $11 to $200 for the last. In Indian services, non-Indian scholars usually pay more, e.g., up to $250 for the first authorship. Less frequently, offers include the option to buy all authorship slots for a paper at once, for a single author or a group of collaborators (Supplementary material, Appendix 6). The price varies from $217 to $750, depending on the service.

Many channels automatically or manually delete offers after a few weeks or months. Thus, we mostly found offers from 2025 since we started searching in 2025, as shown in Figure 2.

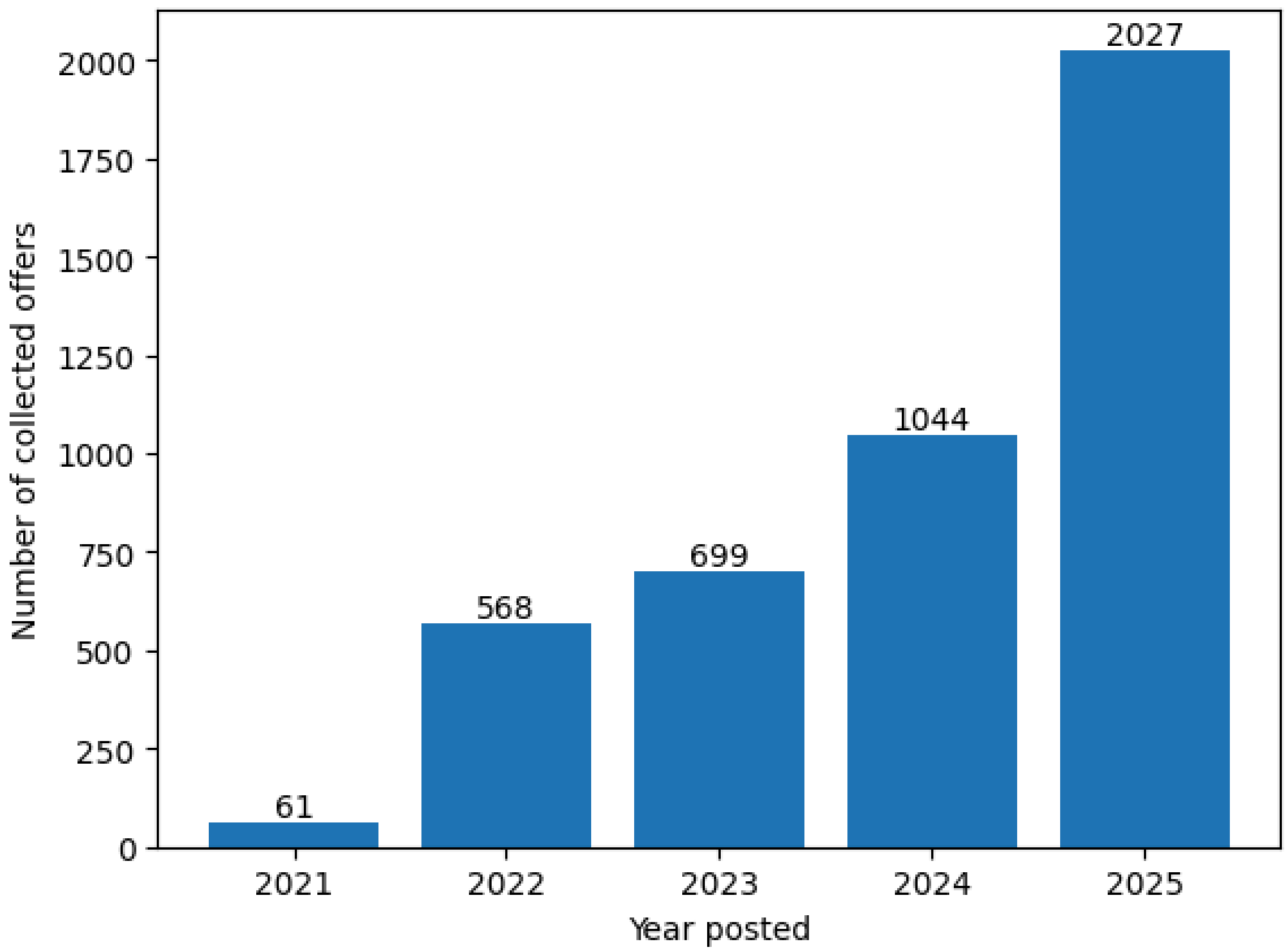

Figure 2. Number of unique collected IEEE offers with an identifiable posting date per year.

## Identification strategy for offers

Our goal was to identify offers whose title is recognizably the same as titles of the final published version, even if it contained minor syntactic changes, so that we could be confident in our matching. We used a combination of manual and semi-automated approaches to match offer titles or keywords to published IEEE conference papers. Manual approaches included searching for the offered titles or keywords in Google Scholar and various web search engines, and using the search functions in IEEE Xplore and Scopus. For a more systematic search, we exported the metadata of all IEEE conference papers published since 2021 from Crossref and, where available, from the conference programs for the latest not-yet-published editions of conferences whose past editions included papers we matched to offers. We then used an automated process to match these titles to the collected offers based on textual or semantic similarity, with subsequent human examination of the most similar pairs. We compared titles using either the Levenshtein distance, which measures the number of characters that need to be added, removed, or substituted to transform one text string to another (we ignored differences in capitalization, punctuation, and spaces), or the cosine similarity between extracted Sentence-BERT embeddings (Reimers & Gurevych, 2019).

Each potential match was independently verified by two co-authors, classifying each paper as “definitely”, “probably”, or “no match”, and we discussed cases in which the two disagreed. We excluded titles for which at least one assessment was “no match” from the final sample, and did additional vetting for the titles assessed as “probably”.

To assess matches, we examined not only the titles, number of authors, and publication dates of the offer and the candidate paper, but also the context for bulk offers. For instance, if multiple titles from the same offer matched papers published at the same conference or by the same co-author, we considered this to be additional proof. For matches rated as “probably”, we also looked for other signs of problematic research practices associated with paper mills, such as plagiarism, collaboration anomalies, tortured phrases (Cabanac et al., 2021), and unrelated references (Clyde, 2025).

We show in Figure 3 an example of conference title offers posted on 1 August 2022, which have been published as part of the same conference held between 14-16 December 2022. The titles of the offered and published papers are identical.

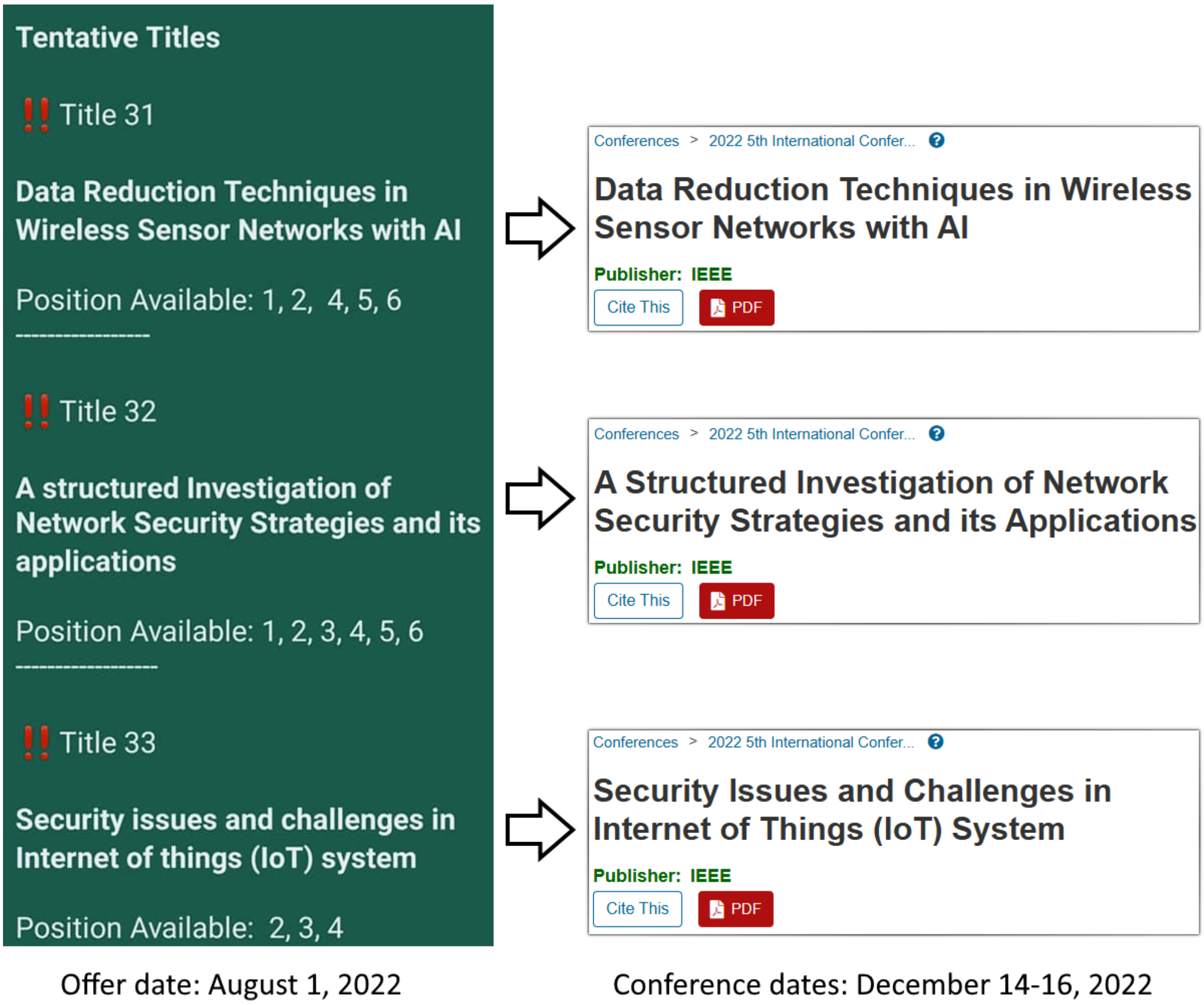

Figure 3: Example of offer identification.

# Results

## Identified papers

We conducted our paper identification process from the start of the offer collection in July 2025 until the end of June 2026. We identified 2,105 unique papers in IEEE conference proceedings that potentially matched the collected authorship-for-sale offers. During our subsequent assessment process, we excluded 43 papers as “no match”, leaving 2,062 papers (Figure 4). Of these, we classified 1,913 as “definitely” matching and 149 as “probably”. In the majority of cases, papers that we classified as “probably” also exhibit signs of likely paper-mill origin; we are merely not absolutely certain if they correspond to the specific advertised offer. 2,051 papers have a DOI and are indexed in the Dimensions database (accessed on July 9, 2026). The remaining 11 have no DOI, thus we could not always include them in subsequent analyses.

The matching of papers to offers was not always a one-to-one assignment. In some cases, we found that the same offer corresponds to more than one published paper, whose contents are often identical or highly similar (e.g., lightly paraphrased) despite having a similar publication date and no authors in common. In a smaller number of cases, a single paper had been advertised both by its full title and, in a separate offer, by only its keywords.

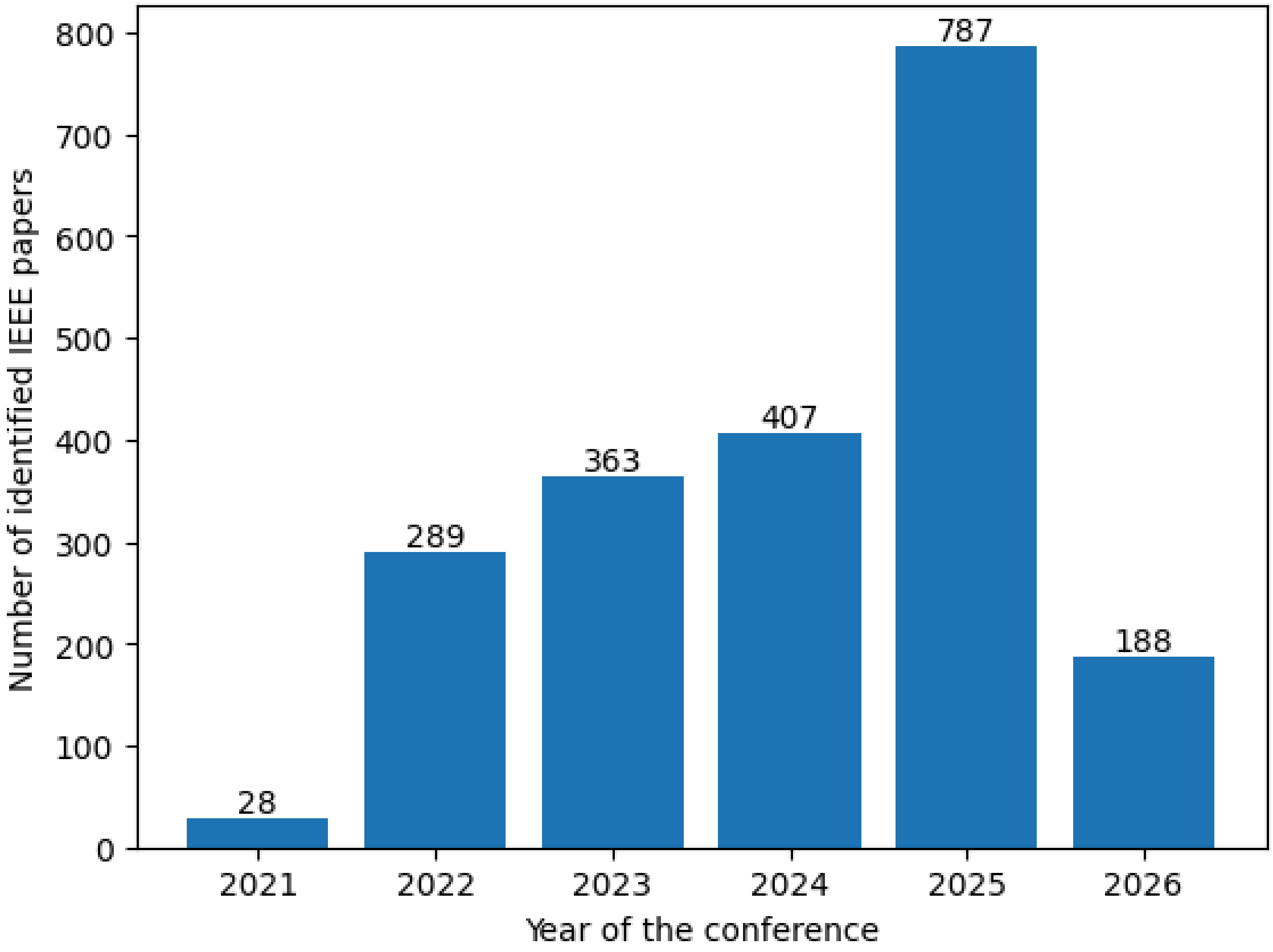

Figure 4: Number of identified IEEE papers per year of the conference

Out of the 2,062 identified papers, 66 have already been retracted for violation of the IEEE authorship policy (38 papers) or for containing tortured phrases (28 papers).

According to the Dimensions database, most of the 2,051 identified papers with DOIs are in the category “Information and Computing Sciences” (81%), with the subcategories “Data Management and Data Science” (27%), “Machine Learning” (15%), “Distributed Computing and Systems Software” (15%) and “Cybersecurity and Privacy” (14%). Other frequent classifications include “Engineering” (18%) and “Commerce, Management, Tourism and Services” (14%). In this classification system, a single paper can belong to more than one category, which is why the percentages sum to more than 100%.

In 1,509 cases of the final sample (73%), the title of the published paper exactly matches the original offer. In many other cases, it is highly similar, e.g., with a slightly different word order or changing a word from singular to plural. In a small number of cases, the paper's published title differs, but the offered title or its parts still appear in the main text.

Even for titles that are not identical, they are often easy to verify because they combine broad computer science terms such as “machine learning” (present in 14% of identified papers) with very specific ones such as “optimal route planning”. Such combinations are very distinctive, often matching a single paper in the entire IEEE catalogue. “Trendy” topics such as “Blockchain”, “Internet of Things” / “IoT”, and “Big Data” are also over-represented in identified titles.

## Authors

According to the Dimensions database, the papers are co-authored by more than 7,500 researchers from over 4,000 affiliations in 56 countries, predominantly from India (93% of papers), the USA (18%), Iraq (11%), Uzbekistan (7%), and Saudi Arabia (4%). The overrepresentation of Indian authors mirrors the country distribution of the collected offers, most of which appear to have been aimed at scholars from India, e.g., listing prices in Indian rupees (INR).

Although most authors appear in only one identified paper, 60 individuals have co-authored at least 10 of these papers each, with the highest number per author being 69 (Figure 5).

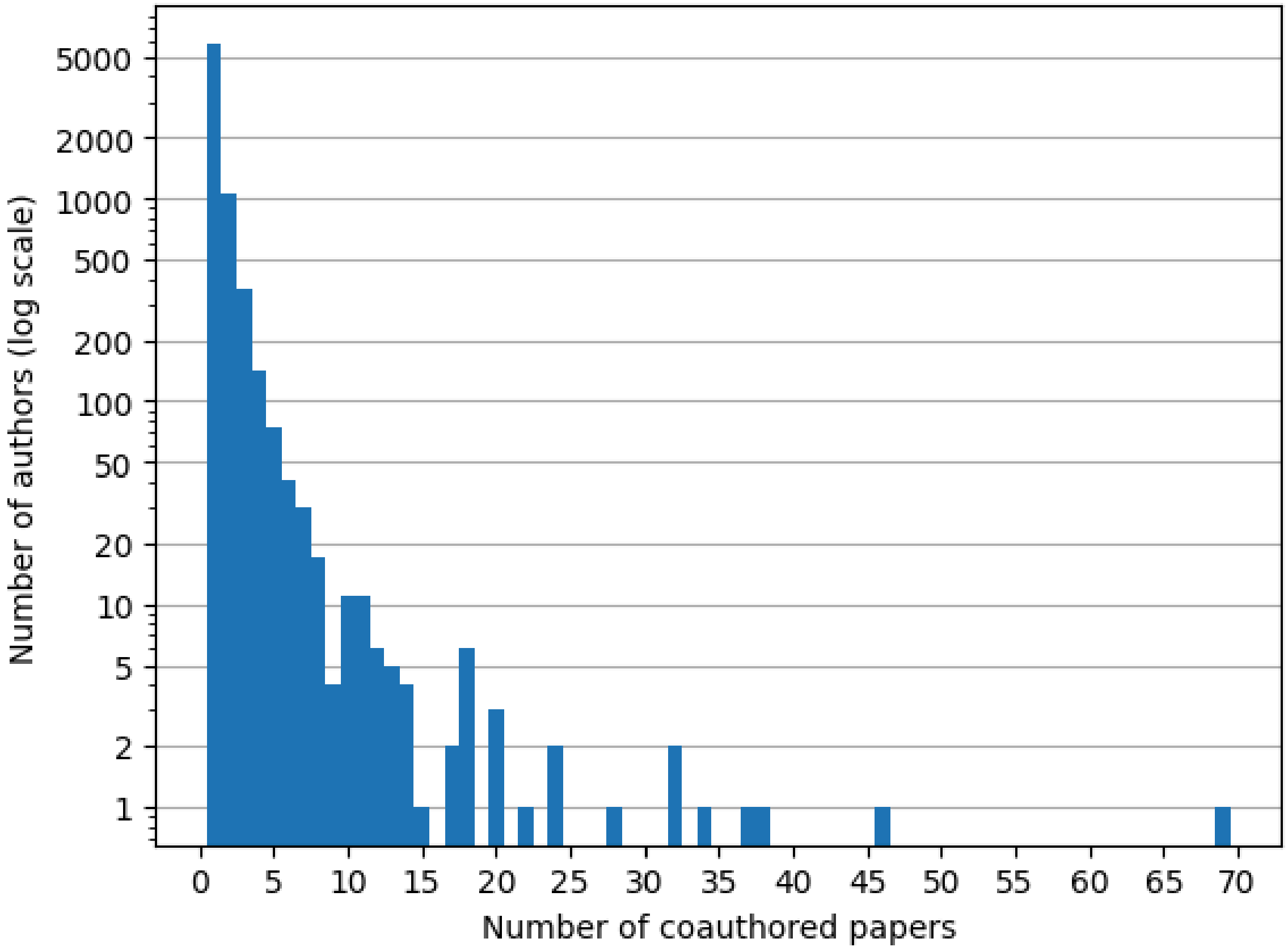

Figure 5: Number of identified publications per author.
Source: Dimensions (accessed July 9, 2026)

## Conferences

The identified papers appear in 366 different conference proceedings published by IEEE. The share of identified papers within individual conference proceedings reaches up to 23.51% (Table 1).

Among the 109 conferences with 5 or more identified papers each, the majority took place in India (95), but some were entirely virtual (4) or held in Uzbekistan (3), Singapore (2), Australia, Indonesia, Malaysia, Nepal, or the United Arab Emirates (1 each).

Most of these conferences were (co-)sponsored by sections of IEEE in India, most frequently by the Uttar Pradesh Section (545 identified papers across 23 conferences), the Madras Section (323 papers across 21 conferences), the Madhya Pradesh Section (162 papers across 13 conferences), or the Bangalore Section (156 papers across 15 conferences). However, society-sponsored conferences are also affected, such as the IEEE Computer Society (34 papers across 2 sponsored conferences, plus an additional 132 papers at 9 conferences co-sponsored by local chapters of the society) or the IEEE Industry Applications Society (43 papers across 3 conferences). For two conferences (ICSES 2022 and ICPCSN 2023), we were not able to confirm the IEEE sponsor, so they are not included in these counts.

This list includes multiple problematic editions of some conference series. For example, we identified problematic papers across three years of the IC3I, ICACITE, and ICONSTEM conferences. This supports the findings of Swart et al. (2025), who observed similar long-term patterns in the number of “tortured articles” in conference series.

Potentially sold authorships are not the only red flag for these conferences. To scan for additional problems, we have consulted the Problematic Paper Screener (Cabanac et al., 2022), an online platform that detects and reports issues in published literature, such as tortured phrases and retracted references. Out of the 50 conferences with DOIs and at least 10 identified papers, 41 have at least one paper identified in the Problematic Paper Screener’s “tortured articles” list (indicating the presence of 5 or more tortured phrases in the paper) and a corresponding PubPeer post, and the remaining 9 have at least one paper identified in the screener’s “clay feet” list, indicating they cite retracted or problematic papers.

Finally, an additional factor that makes many of these conferences suspect is their low submission-to-acceptance time. Out of the 48 conferences with at least 10 identified papers for which we could find archived websites, 13 have a time to acceptance of 10 days or fewer, with the lowest being 3, and an additional 5 have a time to acceptance of 20 days or fewer. Finally, one conference may have dispensed with paper review altogether, as, according to its website, it extended its “full paper submission” date to only one day before the start of the conference itself.

Table 1: Conferences with at least 25 papers matched to authorship-for-sale offers.

| Conference acronym and (IEEE ID number) | Number of identified papers | Papers in the proceedings, % | | |
|---|---|---|---|---|
| | | Identified | Six-author | Six-author with 5+ different affiliations |
| IC3I 2022 (56241) | 95 | 23.51 | 41.09 | 32.18 |
| ICACITE 2023 (57410) | 83 | 14.59 | 51.14 | 43.23 |
| ICONSTEM 2025 (65670) | 63 | 6.32 | 53.11 | 40.41 |
| CE2CT 2025 (64011) | 56 | 21.54 | 45.00 | 32.31 |
| ICRISET 2025 (64803) | 45 | 17.05 | 54.54 | 42.05 |
| ICTBIG 2025 (68706) | 45 | 9.85 | 45.73 | 27.13 |
| ICRTCST 2026 (68392) | 43 | 23.12 | 57.53 | 39.25 |
| MysuruCon 2022 (55714) | 38 | 9.45 | 15.17 | 5.22 |
| ICFTS 2025 (62006) | 35 | 13.46 | 47.31 | 35.77 |
| ICACCM 2024 (61117) | 33 | 12.94 | 50.98 | 36.86 |
| AISC 2023 (56616) | 29 | 10.21 | 22.18 | 13.38 |
| ICTACS 2022 (56270) | 27 | 16.36 | 26.06 | 21.21 |
| ICSES 2024 (63760) | 27 | 6.82 | 55.30 | 32.07 |
| IC3TES 2024 (62412) | 27 | 10.47 | 40.70 | 25.58 |
| ICONSTEM 2024 (60960) | 27 | 8.31 | 39.69 | 24.00 |
| ICICAT 2025 (68430) | 27 | 14.44 | 41.18 | 21.93 |
| SMART 2022 (55829) | 25 | 8.65 | 13.15 | 9.00 |

## Collaboration and citation anomalies

Identified papers show collaboration anomalies that are not expected of typical IEEE papers.

*Number of co-authors.* 82.0% of papers have exactly six co-authors, compared to 15.6% for all conference papers published by IEEE in 2024 (see Figure 6).

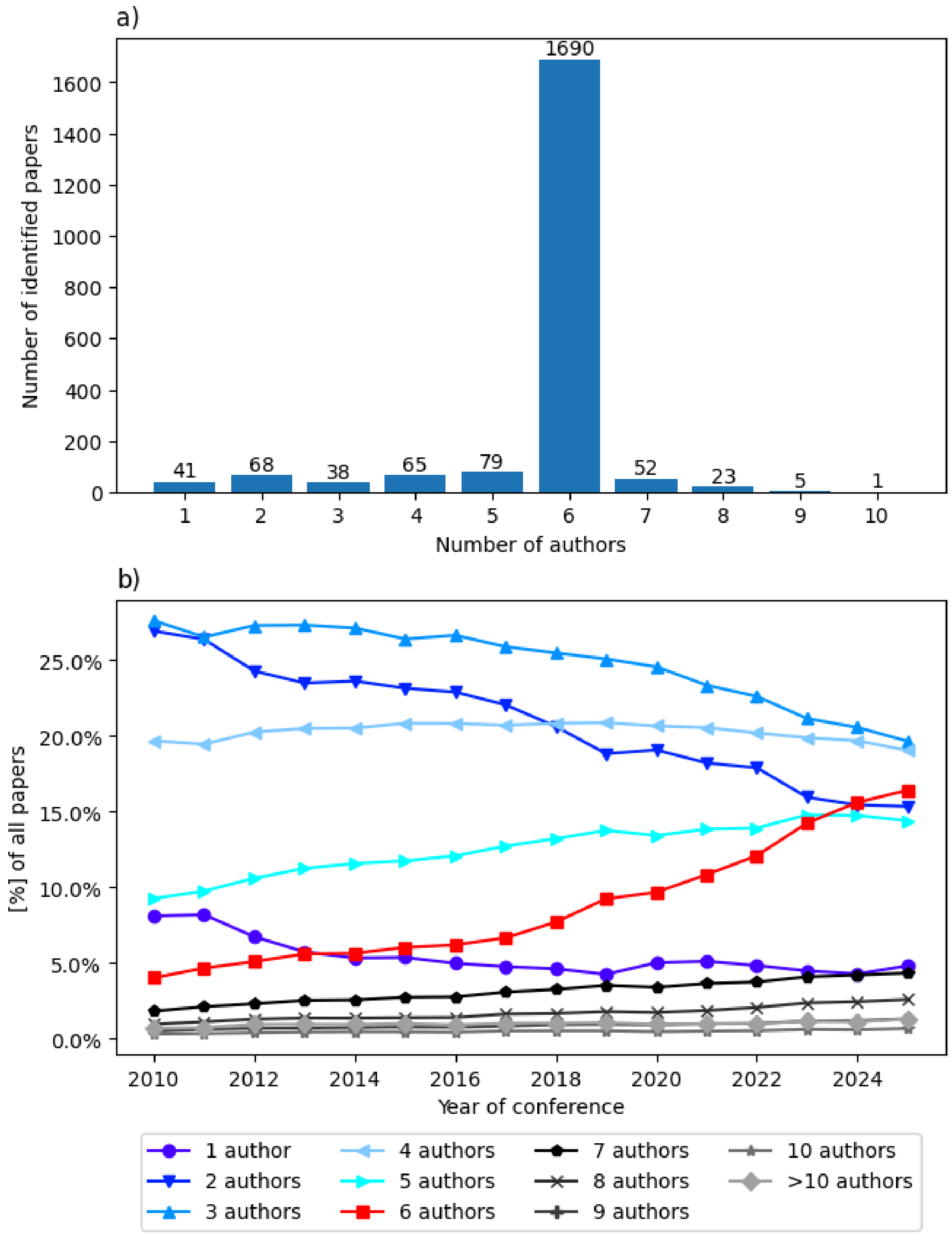

Figure 6: Number of co-authors in a) the 2,062 identified papers, b) all conference papers published by IEEE from 2010 to 2025. Source: Crossref (accessed February 28, 2026).

*Affiliation diversity*. On average, each paper lists 5.1 distinct author affiliations (these can be universities, research institutions, departments within institutions, or companies, as reported by Dimensions). Such a high degree of institutional diversity relative to the number of authors is typical of paper-mill products (Abalkina, 2023). 46% of the identified papers include authors from at least two countries. This suggests that the authorship-for-sale services operate internationally.

*Citations*. Some paper mills offer “citation boosting” services, in which they insert references to a specific author’s publications into their products in order to increase the author’s publication metrics (Clyde, 2025; Ibrahim et al., 2025). We encountered such advertisements during the data collection (Supplementary Materials, Appendix 7). Thus, we analyzed all references in the identified papers for citation irregularities using Dimensions.

As of July 9, 2026, we retrieved the references of 2,038 papers out of 2,062 and identified 34,548 references citing 24,596 unique publications. These references cite 80,303 different authors, with 98.2% of authors being cited fewer than 10 times. 37 authors have each been cited more than 100 times in the identified papers, with the most frequently cited individual receiving 559 citations across 112 citing papers (see Table 2). Among the top-cited individuals, some are co-authors of identified papers, and some are not.

In some cases, references to the top-cited authors were found unrelated to the topic of the citing paper or to the citation context. For example, in the ICACCM 2024 conference, the questionable references were cited in identical off-topic text reused by multiple papers. This suggests that such references were added to increase the citation metrics of specific publications or authors.

Table 2: The 10 authors most cited by the identified papers.

| | **Citations** | **Cited papers** | **Citing papers** | **References per citing paper** | **Co-authored identified papers** |
|---|---:|---:|---:|---:|---:|
| *Author 1* | 559 | 87 | 112 | 4.99 | 69 |
| *Author 2* | 419 | 101 | 192 | 2.18 | 3 |
| *Author 3* | 398 | 52 | 14 | 28.43 | 0 |
| *Author 4* | 382 | 57 | 137 | 2.79 | 2 |
| *Author 5* | 280 | 37 | 103 | 2.72 | 0 |
| *Author 6* | 276 | 56 | 80 | 3.45 | 0 |
| *Author 7* | 273 | 7 | 81 | 3.37 | 0 |
| *Author 8* | 264 | 44 | 107 | 2.47 | 0 |
| *Author 9* | 249 | 71 | 45 | 5.53 | 0 |
| *Author 10* | 243 | 73 | 46 | 5.28 | 22 |

Source: Dimensions.

## Content anomalies

*Superficial content.* Three co-authors with expertise in computer science manually assessed some of the papers and found that many papers in the identified sample feature generic and superficial content, irrelevant examples, and graphs that do not add meaningful information to the paper. More than 50 paper titles also mention one of over 20 different “bio-inspired” algorithm types, for example, “whale optimization algorithm” or “jellyfish search optimizer”. Such algorithms have been the targets of critique due to their abundance and perceived lack of novelty, scientific motivation, and rigor (Aranha et al., 2022; Sörensen, 2015).

*Text similarities*. The identified papers show text reuse in titles, abstracts, and content. At least 13 pairs of papers have identical or nearly identical abstracts. We also observed 46 passages of text that are reused in at least 2 papers. These passages go from a few sentences to nearly the entire content, title included, of different papers. As we described above, some of the duplicate passages contain citations to a large number of papers at once.

*Plagiarism*. Some papers also exhibit high similarity to previously published papers outside the identified sample, including plagiarized or paraphrased passages, figures, and results. Some offered titles appear to have been copied from papers published before the offers.

*Tortured phrases*. According to the Problematic Paper Screener (accessed July 2, 2026), 128 of the identified papers contain at least 5 known tortured phrases each (such as “information the executives” instead of “data management”), which are typically used to disguise plagiarized passages (Swart et al., 2025). The highest number of detected tortured phrases in a single already retracted paper is 47. Among papers that are not retracted at the time of the analysis, the highest number is 36. This analysis excludes 11 identified papers without a DOI, which are not analyzed by the Problematic Paper Screener.

*Emails*. We observed that the identified papers usually use commercial email addresses (e.g., Gmail or Outlook) rather than institutional accounts. Although this is not necessarily problematic, we have found eight papers with the same non-institutional email address for all authors, which is probably associated with an authorship-for-sale service. Similar to this, we noticed another email used by five different scholars across nine papers; the name of the email contained medical terms that didn’t correspond to the topics of the papers.

## Other publishers

Besides the IEEE papers discussed and analyzed above, we also identified 147 papers linked to authorship offers in Springer Nature, Taylor & Francis, AIP Publishing, Elsevier, EDP Sciences, and three other conferences (Table 3). The venue where these papers were published did not always match the original offer, e.g., some titles advertised for IEEE conferences were published in non-IEEE venues, and vice versa. Additionally, 50 of the advertised titles appear to have been published as journal articles or (non-conference) book chapters, despite being originally advertised as conference papers.

Table 3: Total number of collected conference offers and identified papers per publisher*.

| Publisher | Collected offers | Identified papers |
|---|---:|---:|
| ACM | 8 | 0 |
| AIP Publishing | 143 | 22 |
| EDP Sciences | 3 | 11 |
| Elsevier | 56 | 17 |
| IOP | 6 | 0 |
| Springer Nature | 195 | 45 |

| Publisher | Collected offers | Identified papers |
|---|---|---|
| Taylor & Francis | 181 | 49 |
| Other | 133 | 3 |
| Unspecified | 161 | N/A |
| Not conference papers | N/A | 50 |

*- 80 offers listed two possible publishers – IEEE and Springer Nature (65) or IEEE and AIP Publishing (15) – and are counted both here and in Figure 2

# Discussion

Our study shows how traditional authorship-for-sale businesses have evolved into full-service providers, infiltrating the entire research ecosystem. Many paper mills operate publicly and openly advertise their services through hundreds of social media channels and websites. They operate like traditional businesses, developing marketing strategies to attract clients, such as by offering discounts on authorships in packages of several papers or promotions for holidays such as Diwali.

Although we have identified 2,062 published IEEE papers and found at least one match for 45% of the 4,407 collected IEEE offers, the size of the problem is likely larger. Many offers remain undetected because the titles have been renamed or may not have been sold. Our findings probably reveal only the tip of the iceberg. The broader scale is reflected in the authorship patterns of the papers originating from authorship-for-sale services. Our study shows the prevalence of six-author papers in our dataset, which is consistent with existing evidence that a specific number of co-authors can be a pattern of a paper mill (Abalkina & Kleiner, 2025).

Our findings show that the number of six-author papers is growing faster than other types of co-authorship in IEEE conferences and suggest an increasing fraction of paper-mill papers. We observe that since approximately 2019, the number of IEEE-published conferences with a larger-than-usual proportion of six-author papers has increased (Figure 7). This phenomenon is not seen with other author counts, which remain more consistent over time. Though this pattern is not a sign of misbehavior, it may indicate that paper mills have infiltrated many conferences.

Table 1 provides further support for a larger-scale problem. For conferences with the highest number of identified papers, we also present the overall share of papers with exactly six authors and at least five different affiliations. In many cases, these numbers are suspiciously high, reaching up to 43.23% of an entire conference proceedings, and far surpassing the number of papers we have identified. Such diversity of affiliations per paper means that they are probably sold to unrelated co-authors who might not have prior collaboration, and is very typical of papers with sold co-authorship slots. Such collaboration anomalies and findings are consistent with previous studies. Abalkina (2023) examined collaboration anomalies in papers potentially originating from a Russian-based paper mill, including diversity of affiliations and mismatches between co-authors' disciplines and the topic of the paper. Guba (2025) found that questionable journals are characterized by larger collaboration teams, which may be a result of an artificial and transactional approach to authorship rather than legitimate teamwork.

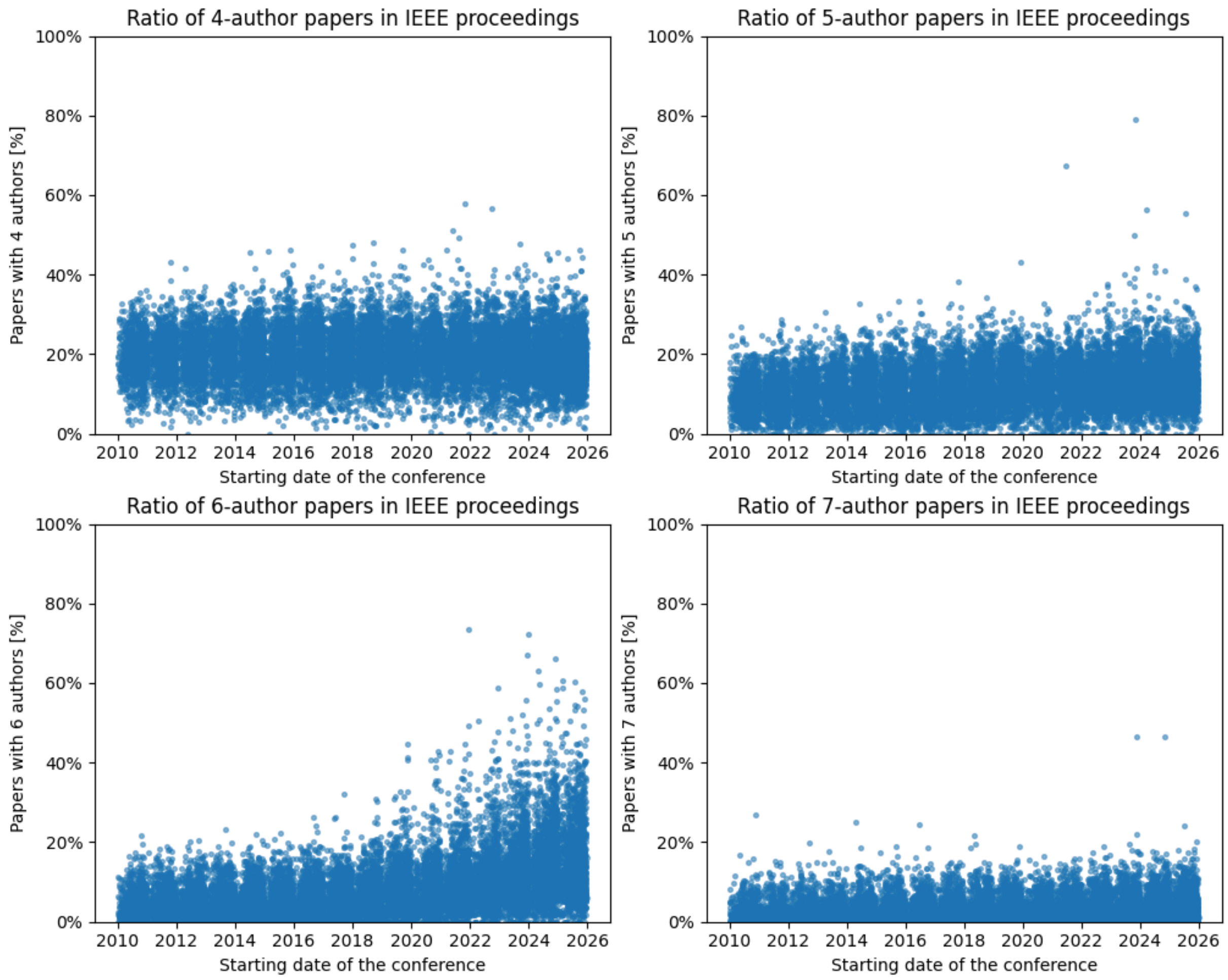

Figure 7: Ratios of four-, five-, six-, and seven-author papers in conference proceedings published by IEEE. Each point represents a single conference with at least 50 papers in its proceedings, held during 2010-2025 and added to IEEE Xplore before March 1, 2026. Plots for other author counts (1–3, 8–11+) are in the Supplementary materials (Appendix 8). Source: Crossref.

International authorship in such papers does not signal increasing international collaborations but rather reflects the market, where paper mills can reach clients from all over the world through social media. Guba (2025) similarly found that in questionable journals, international collaboration happens between countries that face similar publication pressure. Papers become not a result of scientific work but a platform for purchased authorship.

As paper mills currently proliferate, collaboration anomalies should be taken into consideration when making scientometric analyses because such co-authorship may bias the analysis of co-authorship patterns, international collaborations, and individual author scores.

Citation manipulation is another type of business of paper mills that accommodates clients who purchase co-authorship slots and other scholars. Citations are used for hiring and promotion (Baccini et al., 2019), which creates incentives for questionable practices such as the increase of self-citations (Baccini & Petrovich, 2023; Seeber et al., 2019) or citation manipulations. Recent studies provide evidence that citations can be artificially increased by individual researchers, citation cartels, and paper mills (Antkare, 2020; Clyde, 2025; Ibrahim et al., 2025), suggesting that citation-based research evaluation is vulnerable to manipulation.

One of the explanations why scholars apply to authorship-for-sale services is the “publish or perish” culture. In our sample, 22.1% of the papers include co-authors affiliated with one specific university in India, which is associated with publication requirements for PhD defense and promotion. Two other top universities, appearing in 14.3% and 13.4% of papers, have similar research evaluation requirements.

The decentralized nature of many conferences may make them especially vulnerable to paper mills. In the case of IEEE, local sections organize conferences and publish their proceedings with no apparent checks or balances. Similar to guest-edited special issues in journals, many conference publishers, including IEEE, are not directly involved in the paper submission, review, and revision process. They only receive the final “camera-ready” versions of papers from the conference organizing committee. As a consequence, they are not aware of potential authorship changes that occur after acceptance. Despite growing awareness of paper mills and numerous available tools to detect some milled papers, recent conference proceedings still exhibit tortured phrases, plagiarism, and manipulated citations.

IEEE’s stance on conferences may be another reason why it is so vulnerable. Despite conferences being considered equivalent or even more important than journals in many subfields of computer science, they are relegated to “Other Publications” in IEEE’s operations manual (IEEE, 2026, §1.5.2) and not considered as publishing “fully developed” work (id., §8.1.7.F). Whereas IEEE considers a journal article as the “primary means” of publication for work with “lasting value” (id., §8.2.5.A), it only requires a conference article to be “not so poorly written that it is unreadable” (id., §8.2.2.B.2). Despite this, IEEE publishes conferences that are at the very top of their field, such as IEEE CVPR being the 2nd-highest-ranked venue by H-index according to Google Scholar at the time of writing.

More generally, conferences are viewed in most subfields of science as lesser than journals, yet many conferences have their proceedings published by prestigious publishers and indexed in Scopus, which means conference papers count towards publication metrics. This mismatch between how important a venue is for authors and how seriously it is taken by its organizers and publisher makes the job of paper mills easier.

## Call to action

We propose four concrete actions that IEEE and other conference publishers should take to fight authorship-for-sale services and prevent further damage to the reliability of the scientific record.

First, screen submissions for red flags such as those we identified in this study: unusually many papers with the same number of authors, a high diversity of affiliations, text reuse, and tortured phrases. This requires tracking papers for their entire lifecycle and not only their final “camera-ready” versions. The recently announced partnership between IEEE and the Clear Skies integrity service (Day, 2026) gives us some hope in this regard.

Second, proactively monitor social media to find services that violate scientific integrity, such as authorship-for-sale services, adapting the red flags from the previous point based on empirical evidence, since services are likely to change their methods to avoid detection.

Third, develop an expedited investigation and systematic retraction process for organized manipulation of the peer-review process and paper-mill activities, as suggested by COPE guidelines (COPE, 2025). The kind of problems we highlight in this study can be spotted quickly, but happen too often for lengthy single-paper investigations to be effective.

Fourth, consider conference-level investigations and retractions of the whole conference with a significant number of problematic papers originating from paper mills. In such cases, they compromise the whole conference proceedings, and conference-wide action, including its retraction, may be applied. Recurring compromised conferences should be discontinued due to their repeated association with paper mills and systematic activities to publish problematic papers from paper mills.

While these recommendations require additional effort on the part of the publishers, many of the necessary tools already exist; all that remains is to dedicate enough resources to the task. As the number of publications and the scale of the paper mill problem increase, publishers must also scale up their research integrity teams; otherwise, they will soon become overrun.

We sent a draft version of this manuscript to IEEE’s research integrity team for feedback. Although we did not receive any comments before submission, we are in communication with them about our findings.

## Limitations

While we found a significantly larger number of sold papers than we originally expected, our data collection still has significant limitations regarding language and timeline.

First, we mainly collected offers in English from social media platforms that are popular in the English-speaking world. We did not, for instance, search through the Chinese equivalents of WhatsApp or Facebook, given that none of us understands Chinese.

Our data is thus likely to be biased towards English-speaking countries, of which India is the largest. While there is evidence that India is home to many paper mills (Basava, 2025; Richardson et al., 2025a), and we found offers from many Indian entities that appear to be independent of each other, this study should not be interpreted as evidence that other countries are immune to it. For instance, while China is also known to be associated with many paper mills (Hvistendahl, 2013; Schneider & Clyde, 2020), we found none, presumably due to our choice of social media platforms.

Second, no single study can identify all channels that sell co-authorships in conference papers, and even for the channels we identified, we found evidence during our collection process that paper mills change paper titles for publication, delete already sold offers, and even delete entire channels. Some channels were removed while we were collecting offers, meaning there could be offers that had been deleted before we even started looking. Therefore, paper mills may be more widespread in conferences than our data suggests.

Finally, our list of identified papers includes only publications available online by June 30, 2026, the date when we finalized our list of potential matches to focus on their assessment and analysis. However, since the publication process takes time, more papers matching the collected offers are still being published, with at least 28 new potential matches appearing in the first 15 days of July 2026. As such, it is certain that the number of published papers matching the collected offers will be even higher than the 2,062 papers we report in this work.

# Conclusion

This research estimated the penetration of paper mills in conference proceedings. We introduced a semi-automated method to identify papers whose authorship was offered for sale. We detected 2,062 papers that are potentially linked to authorship-for-sale operations and published in 366 IEEE conferences, and 147 papers linked to conferences of other publishers. The real size of the problem is likely to be significantly larger.

The identified papers demonstrate collaboration and content-based anomalies, including an abnormal diversity of affiliations, a predominance of six-author papers, citation manipulation, plagiarism, and text reuse across papers of the sample, tortured phrases, and superficial content. These "red flags" may help publishers to screen and identify problematic papers.

Our findings advance the understanding of authorship-for-sale mechanisms and paper mill ecosystems across social media platforms. Offers for sale are publicly accessible and use direct language to sell papers. We identified over 200 channels that (re-)post offers of conference papers for sale with a cumulative audience of more than 600,000 people. Such businesses extend their operations to a wider range of services, including citation boosting, patents, books, editorial positions, etc. Paper mills commercialize misconduct in different aspects of the research environment and create an organized and scalable market to promote dishonest services.

Like Pandora's box once opened, looking into paper mill proliferation in conferences reveals fraud, misconduct, and authorship for sale that can no longer be ignored. What remains is hope that scrutiny, professional editorial work, detection systems, prompt corrections, and collective action can protect the scientific literature.

# Acknowledgements

We would like to thank Dr. Nick H. Wise for providing additional data from his own collection of authorship-for-sale offers. We are also grateful to the sleuth community on PubPeer, whose pre-existing comments on some of the identified papers were helpful during our assessment. We thank Dimensions for providing access to their database. We are grateful to Adrian Barnett, Shaoxiong Brian Xu, and an anonymous reviewer for their valuable comments on an earlier version of this manuscript.

# Funding

YO acknowledges the NanoBubbles project that has received Synergy grant funding from the European Research Council (ERC), within the European Union's Horizon 2020 program, grant agreement no. 951393 (https://doi.org/10.3030/951393). Open Access funding enabled and organized by Projekt DEAL.

# Data availability

Our collected data, which includes offers, matched papers, channels, conferences, offer screenshots, as well as some scripts used for data analysis, can be found at https://doi.org/10.5281/zenodo.19727148.

# Conflicts of Interest

AA gave an invited talk at an IEEE conference in 2024 and was reimbursed for travel and accommodation costs. At the time of submission, MK is a member of the IEEE Signal Processing Society.

# Use of Gen AI

The authors used Gen AI for grammar checking only, not for editing or writing the text. The authors used OCR technology, including Gen AI models, to identify the text in images, and the results were manually checked.

## SUPPLEMENTARY MATERIALS

APPENDIX 1

Presentation for a conference

Today

[redacted]

Dear Professors,
We have received **acceptance** for two conference presentations through our company [redacted] scheduled for 28.11.2025 & 29.11.2025
We have also prepared the PPT for the sessions.

May I kindly check if you would like to present, or should we proceed with the presentation from our side?
Please let us know at your earliest convenience so we can finalize the presenter details.

Looking forward to your confirmation.

**IEEE Title id**

[redacted]

Warm regards,
[redacted]

09:28

APPENDIX 2

Other services provided by co-authorship-for-sale services

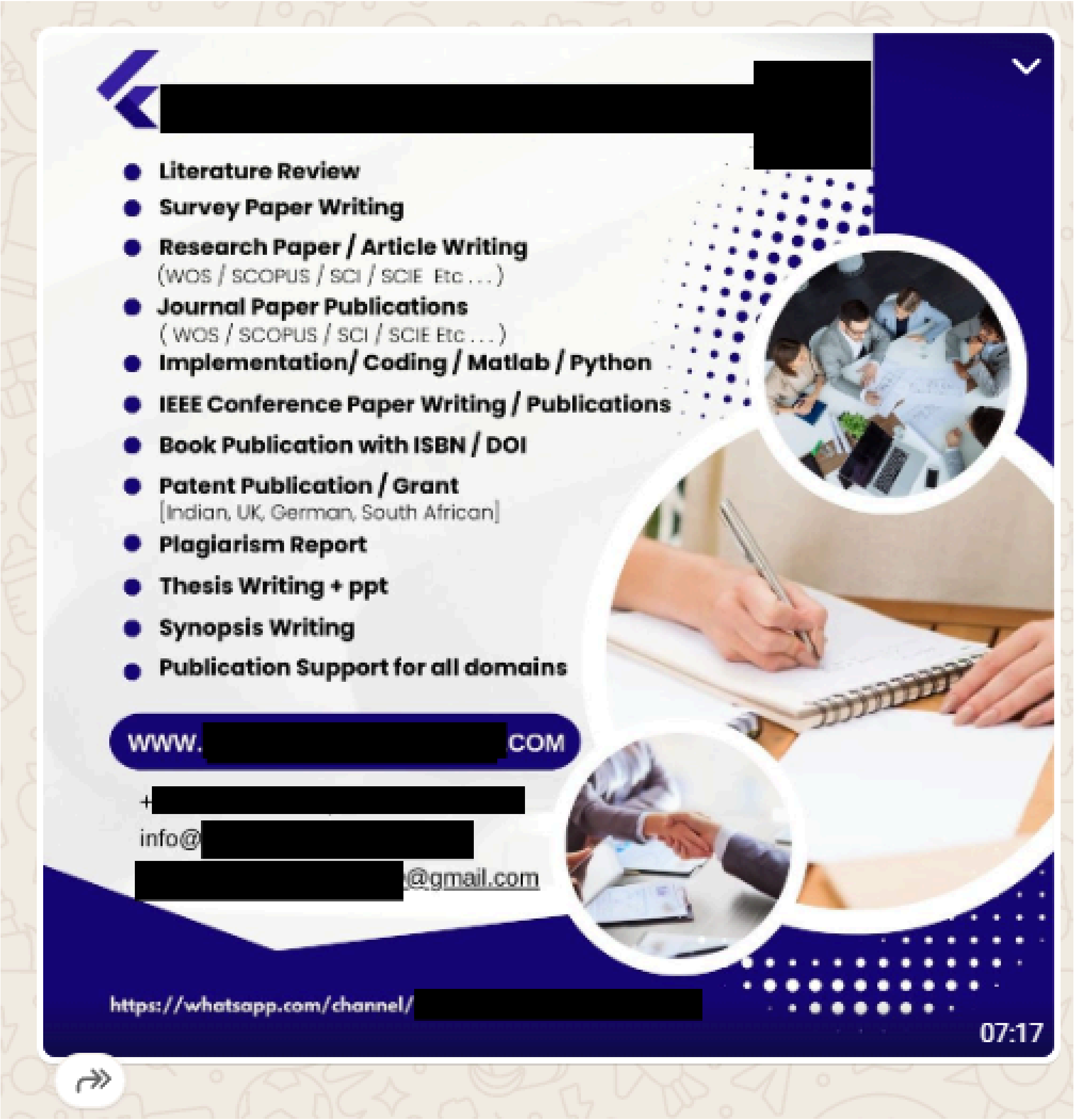

8/19/2025

[redacted]

Community admin

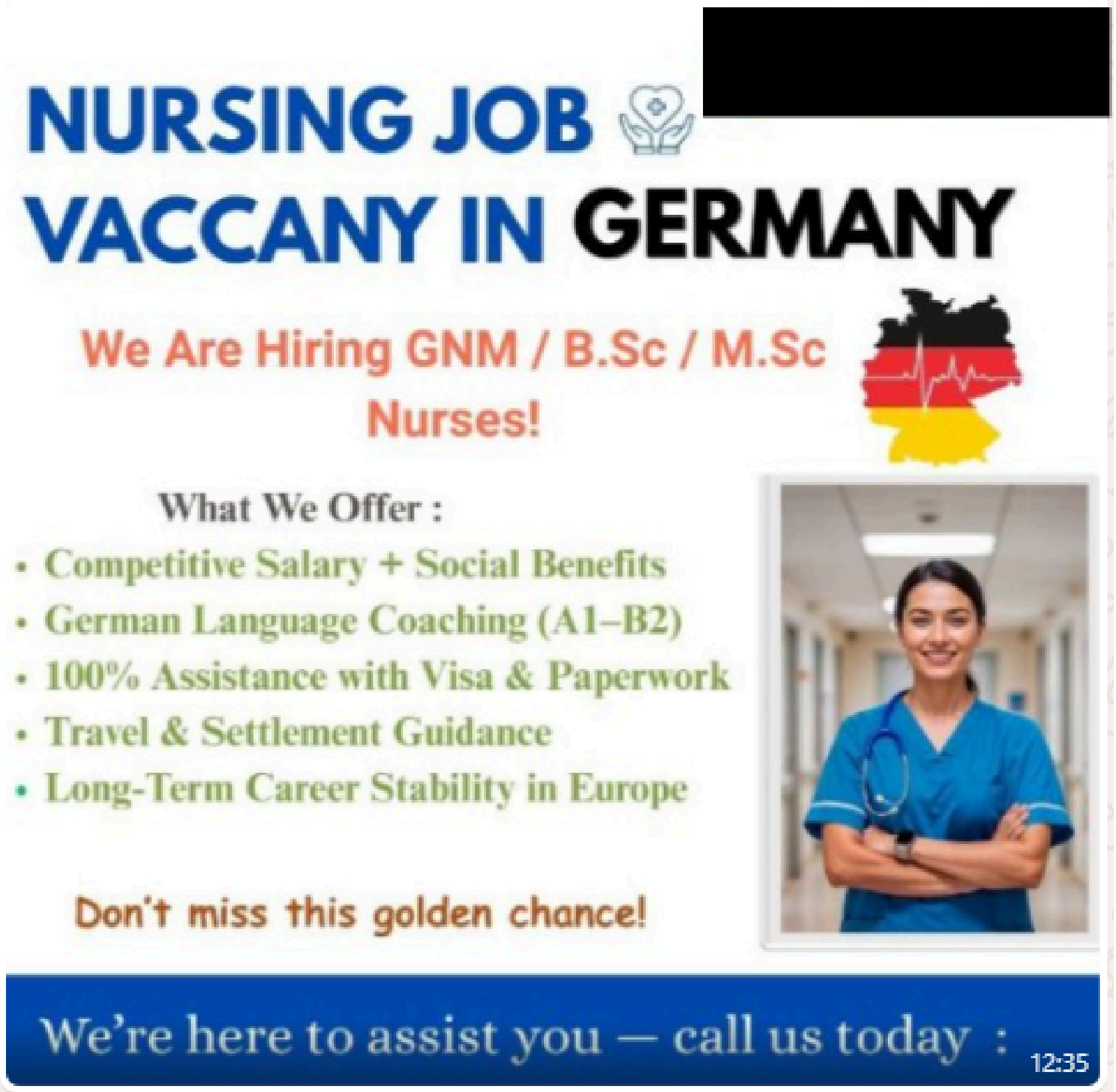

[redacted]
Community admin
Forwarded
🌟Call for AWARDS 2025🏆📚

A fantastic opportunity to enhance your profile felicitate the Higher Educational Institutions, Academicians, Research Scholars, and Young Engineers through the precious Award 🎖️

Focus will be given to Academicians, Researchers, Scientists and Professionals from Industries👩‍🎓👨‍🎓

Benefits of the Awards

1. Brass Type Memento 🎖️

2. Certificate with respective Award Title📜 (Soft Copy/ Hard Copy)

3. All the Awardees Details along with photo will be published in Website🖥️

4. Professional Membership with ID card👨‍💼👨‍💼

5. Fast Processing Time 📮

💥 AWARD CATEGORY 💥

1. Emerging College of the Year Award
2. Principal of the Year Award
3. HOD of the Year Award
4. Best Academic Department
5. Best Academician Award
6. Research Excellence Award (Above 35 Years)
7. Young Researcher Award (Below 35 Years)
8. Young Achiever Award (Below 35 Years)
9. Best Teacher Award
10. Academic Excellence Award
11. Best IQAC Coordinator
12. Best NAAC Coordinator
13. Best ISO Coordinator

APPENDIX 3

Offer in an official IEEE group on Facebook

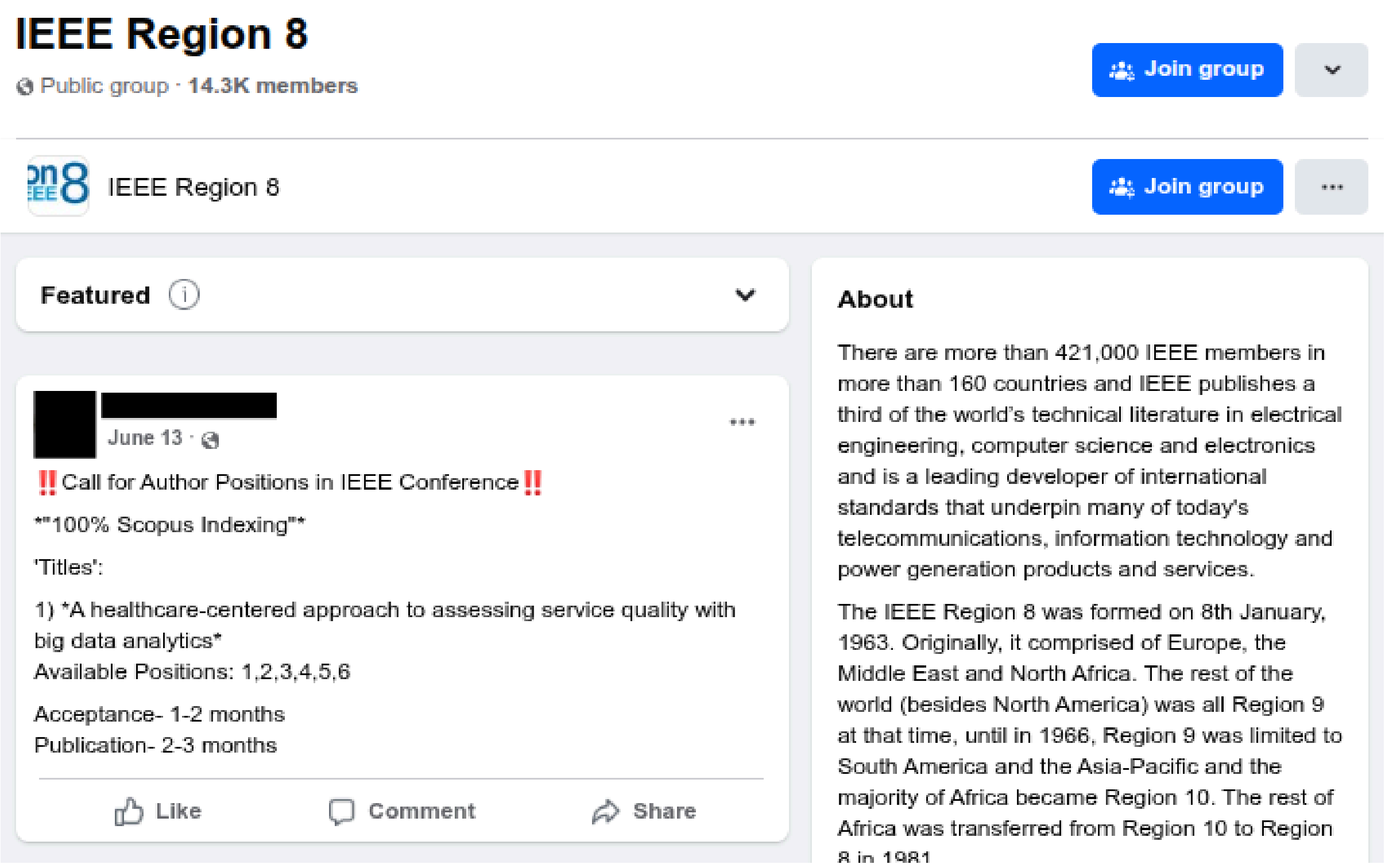

APPENDIX 4

Selling already accepted / already presented papers:

[redacted]
July 25, 2024 ·

Calling for *Springer conference- 100% index in Scopus*

*Already accepted and Conference also over*
Conference done in reputed NIT

Only 4 authors in each title

Title 1:
Monitoring and Optimize the energy management of microgrid
Available Positions
1,2,3,4

Title 2:
Smart electrical panner monitoring system
Available Positions
1,2,3,4

Title 3
Experimental forecasting to predict the thermal energy storage system under different load conditions
Available Positions
1,2,3,4

Title 4:
MIoT smart home energy monitoring system
Available Positions
1,2,3,4

Wa.me/[redacted]

APPENDIX 5

Examples of offers with 6 positions

**Publisher: Springer**
**Duration: 3-6 Months,**
**Domain: CSE/ECE/EEE/IT/BIO/MBA/Other depts,**
**Maximum: 6 Authors only.**

Springer

[redacted]

**Note: Only 50% Advance, Remaining After Acceptance of Paper ( +[redacted] )**

| sno. | Title | POSITION : 1 | POSITION : 2 | POSITION : 3 | POSITION : 4 | POSITION : 5 | POSITION : 6 |
|---|---|---|---|---|---|---|---|
| 1 | AI-Driven Personalized Healthcare: Integrating Federated Learning for Secure Patient Data Processing | AVAILABLE | AVAILABLE | AVAILABLE | AVAILABLE | AVAILABLE | AVAILABLE |
| 2 | Large Language Models in Banking: Enhancing Fraud Detection and Risk Assessment Using NLP | AVAILABLE | AVAILABLE | AVAILABLE | AVAILABLE | AVAILABLE | AVAILABLE |
| 3 | Cybersecurity in the IoT Era: AI-Powered Intrusion Detection and Anomaly Detection Systems | AVAILABLE | AVAILABLE | AVAILABLE | AVAILABLE | AVAILABLE | AVAILABLE |
| 4 | Optimizing Software Testing with AI: A Reinforcement Learning Approach for Automated Bug Detection | AVAILABLE | AVAILABLE | AVAILABLE | AVAILABLE | AVAILABLE | AVAILABLE |
| 5 | Explainable AI for Healthcare: Bridging the Gap Between Black-Box Models and Clinical Decision-Making | AVAILABLE | AVAILABLE | AVAILABLE | AVAILABLE | AVAILABLE | AVAILABLE |
| 6 | A Deep Learning-Based Framework for Real-Time Cyber Threat Intelligence and Mitigation | AVAILABLE | AVAILABLE | AVAILABLE | AVAILABLE | AVAILABLE | AVAILABLE |
| 7 | Blockchain-Integrated AI for Secure Data Governance in Financial Services | AVAILABLE | AVAILABLE | AVAILABLE | AVAILABLE | AVAILABLE | AVAILABLE |
| 8 | Edge AI in IoT: Enhancing Real-Time Data Processing for Smart Healthcare Applications | AVAILABLE | AVAILABLE | AVAILABLE | AVAILABLE | AVAILABLE | AVAILABLE |
| 9 | Automated Software Quality Assurance: AI-Powered Defect Prediction and Code Optimization | AVAILABLE | AVAILABLE | AVAILABLE | AVAILABLE | AVAILABLE | AVAILABLE |
| 10 | Adversarial Attacks on Large Language Models: Assessing Vulnerabilities and Defense Mechanisms | AVAILABLE | AVAILABLE | AVAILABLE | AVAILABLE | AVAILABLE | AVAILABLE |

[redacted]

| S.No | **IEEE Conference Proceedings - Paper Title** Acceptance 3 months Publication 4 months After Conference \| **Scopus Index** | **1st Position** | **2nd Position** | **3rd Position** | **4th Position** | **5th Position** | **6th Position** |
|---|---|---|---|---|---|---|---|
| 1 | Dynamic Modeling and Control of Robotic Manipulators Using Reinforcement Learning | Booked | Booked | Booked | Booked | INR 5500 | INR 5000 |
| 2 | The Impact of Social Media Algorithms on Public Opinion Formation and Polarization | Booked | Booked | Booked | INR 6000 | Booked | Booked |
| 3 | Resilient Infrastructure Design: Integrating AI for Enhanced Disaster Preparedness and Response | Booked | Booked | Booked | INR 6000 | INR 5500 | Booked |
| 4 | Solutions: Transformative Strategies for India's Pharma Industry by 2030 | Booked | Booked | Booked | Booked | INR 5500 | INR 5000 |
| 5 | Customer Behavior Forecasting Using Cloud-Based Machine Learning Models | INR 7500 | Booked | Booked | Booked | Booked | INR 5000 |
| 6 | Artifical Intelligence in predictive Maintenance : A Machine learning Approach for Smart Factories | Booked | INR 7000 | Booked | INR 6000 | INR 5500 | INR 5000 |
| 7 | Intergrating AI with Robotics for Enhanced Industrial Automation | Booked | Booked | INR 6500 | Booked | INR 5500 | INR 5000 |
| 8 | Cloud-Based Real Time Traffic Management Systems for smart cities | Booked | INR 7000 | INR 6500 | Booked | INR 5500 | INR 5000 |
| 9 | Advancements and Emerging Trends in Electrical Machines Enhanced by Artificial Intelligence | Booked | INR 7000 | INR 6500 | INR 6000 | INR 5500 | Booked |
| 10 | Sustainable Engineering Practices: Implementing Green Technologies in Industrial Management | INR 7500 | Booked | Booked | INR 6000 | INR 5500 | INR 5000 |
| 11 | Leveraging Data Analytics for Enhancing Guest Experience: Strategies and Impacts in Modern Hotel Management | Booked | Booked | INR 6500 | INR 6000 | INR 5500 | Booked |
| 12 | Leveraging IoT for Real-Time Monitoring of Food Safety and Quality in Supply Chains | Booked | INR 7000 | INR 6500 | INR 6000 | INR 5500 | Booked |

[redacted] + Follow

3mo •

Connect with us on WhatsApp: http://wa.me/[redacted]

Email us : info@[redacted]
Visit Our Website: www.[redacted]
Join our group:
https://lnkd.in/[redacted]

#publicationservices #writingservices #Patent #thesiswrit
#Elsevier#IGI book chapter
publication #Plagiarism checking & reduction #

1

Like Comment Repost Send

Add a comment...

June 3

**IEEE Conference publication**

IEEE conference accepted papers co-author position available
interested DM me

32 1:27 PM

**IEEE Conference publication**

*Author Position available for IEEE conference Publication.*

*Title of the paper: Implementation of a deep learning model for Realtime detection of Diabetic retinopathy in primary care clinics *

Available Position details,

1st Author Position: Booked
*2nd Author Position : Available
*3rd Author Position : Available
4th Author Position: Booked
5th Author Position: Booked
6th Author Position: Booked

33 1:31 PM

**IEEE Conference publication**

*Author Position available for IEEE conference Publication.*

*Title of the paper: Realtime AI driven Monitoring of ICU patients for early detection of Acute Respiratory Distress Syndrome *

Available Position details,

1st Author Position: Booked
*2nd Author Position : Available
*3rd Author Position : Available
4th Author Position: Booked
5th Author Position: Booked
6th Author Position: Booked

38 1:34 PM

# APPENDIX 6

All authorship slots being sold at once, with higher prices for non-Indian scholars

📢 Call for Papers – IEEE Conference Proceedings (100% Scopus Indexed)
🎯 Full Article Titles Available | Guaranteed Scopus Indexing

Today

We are pleased to announce that full article titles are now available for upcoming IEEE Conference Proceedings. This is an excellent opportunity for researchers, scholars, and academicians to publish their original work in 100% Scopus-indexed IEEE conference proceedings.
📝 IEEE Writing & Publication Charges :

₹36,000 – For Indian Authors

₹50,000 – For Foreign Authors ✅ Available Full Article Titles:

Leveraging Machine Learning and Deep Learning to Optimize Food Processing and Nutritional Analysis in Food Technology

Machine Learning Approaches for Optimizing Thermal Performance in High-Performance Mechanical Systems: A Data-Centric Analysis

Deep Learning-Powered Drug Discovery: Enhancing Pharmacognosy and Pharmaceutics Through AI-Driven Molecular Analysis and Prediction

A Deep Learning Approach for Predictive Modeling of Microbial Behavior in Space-Based Bioreactors

Deep Learning Approaches for Predictive Handover and Mobility Management in Ultra-Dense 5G Network Environments
📌 Limited Slots Available!
📞 For Booking & Details:
Mobile / WhatsApp: [redacted]

08:48

APPENDIX 7

Citation boosting

[redacted] [redacted]

10/16/2025

[redacted]

Dr.[redacted]

Professor,[redacted]

scholar.google.com

🥳🥳🥳 **H-INDEX BOOSTING OFFERS** 🥳🥳🥳

**GOOGLE SCHOLAR Citation Package**

**100 Citations/ 10k**

https://scholar.google.com/citations?user=[redacted]&hl=en

**Scopus Citations: ₹ 250 per citation**

Contact [redacted]/
[redacted]

18:05

APPENDIX 8

**Ratios of N-author papers in individual conference proceedings published by IEEE, for (a) N=1,2,3 and (b) N=8,9,10 and N>=11 (the plots for N=4,5,6,7 are shown in Figure 7). Each point represents a single conference with at least 50 papers in its proceedings, held during the years 2010-2025 and added to IEEE Xplore before March 1, 2026. Source of the data: Crossref.**

**a)**

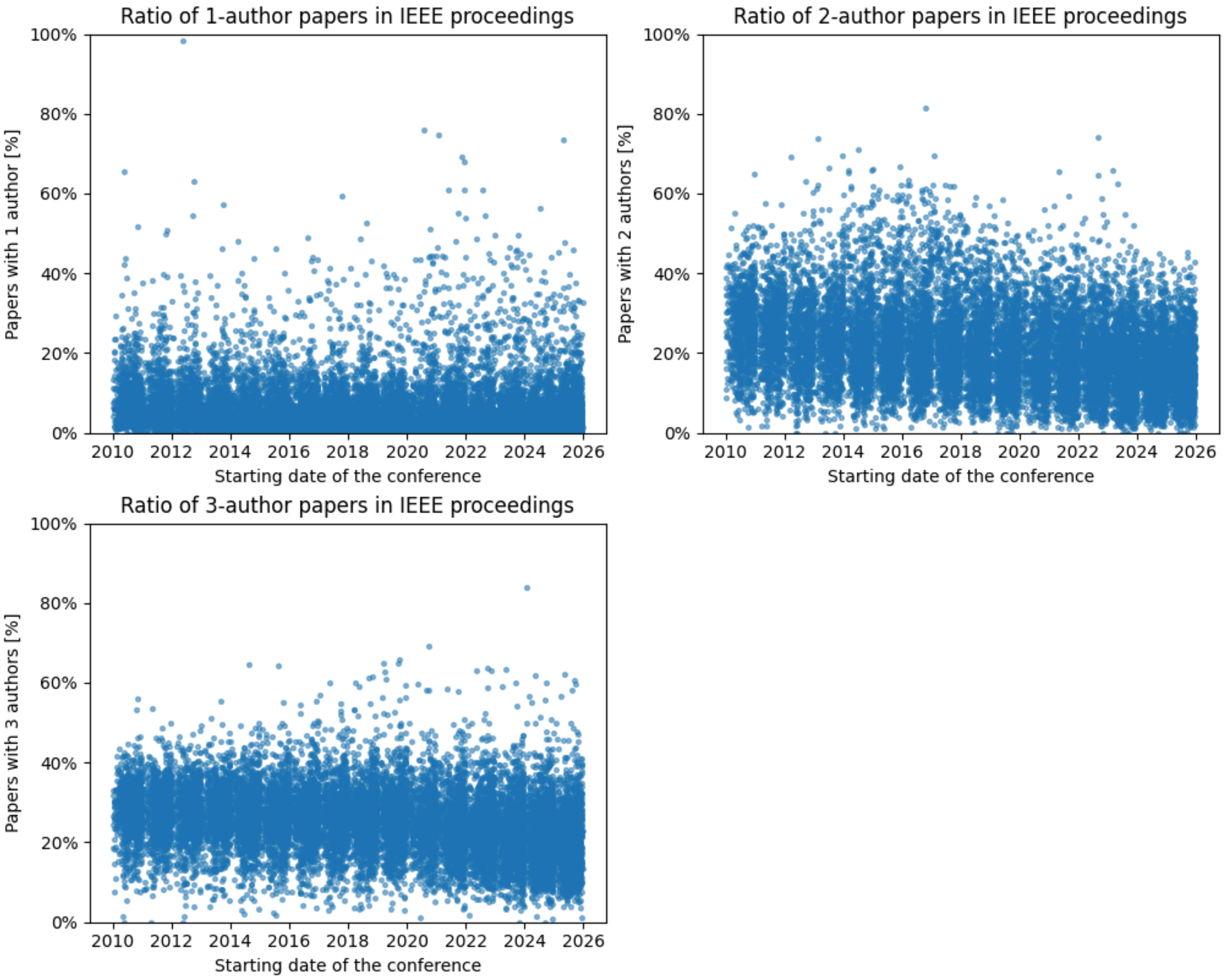

b)

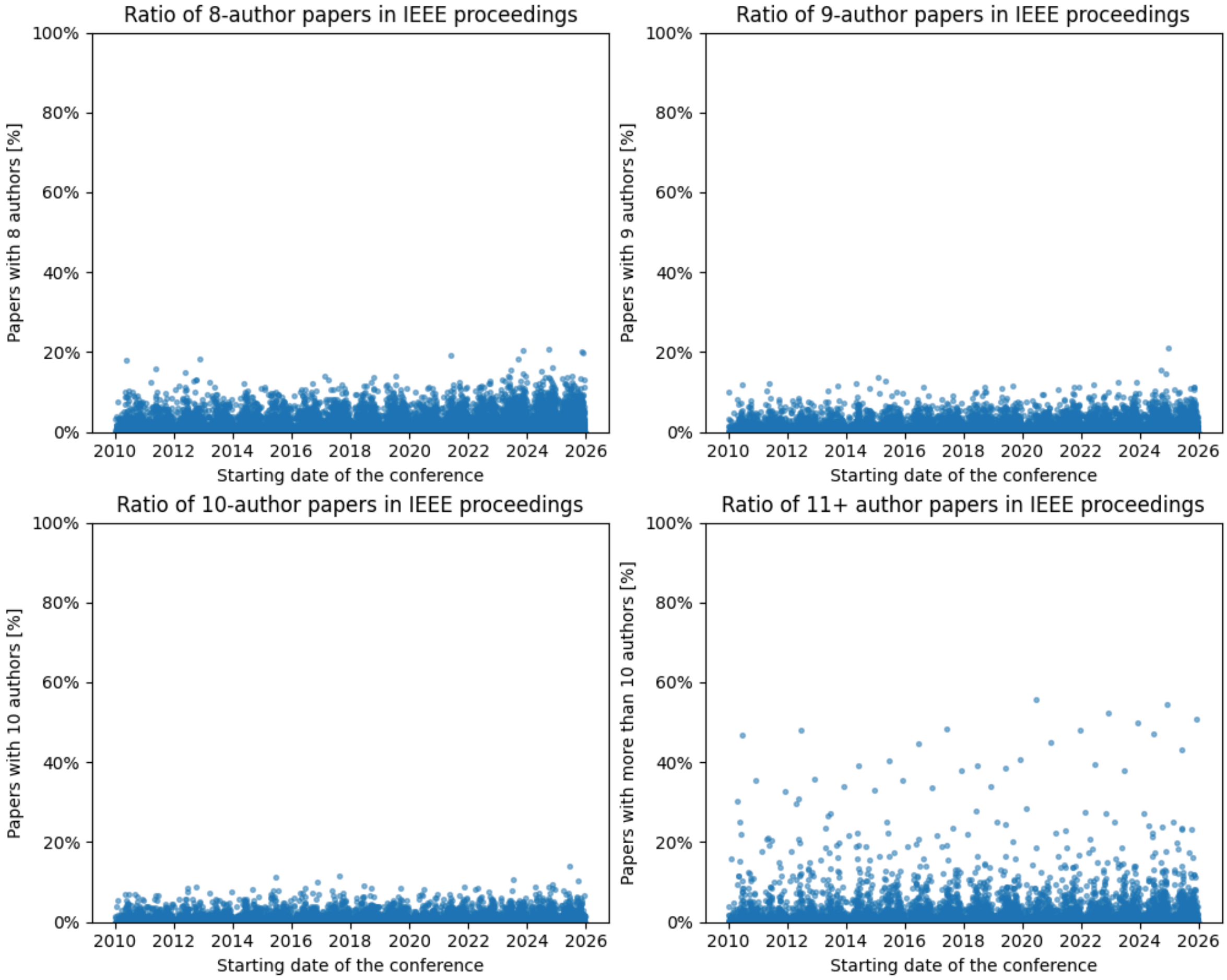
Ratio of 8-author papers in IEEE proceedings
Papers with 8 authors [%]
Ratio of 9-author papers in IEEE proceedings
Papers with 9 authors [%]
Ratio of 10-author papers in IEEE proceedings
Papers with 10 authors [%]
Ratio of 11+ author papers in IEEE proceedings
Papers with more than 10 authors [%]
100%
80%
60%
40%
20%
0%
2010
2012
2014
2016
2018
2020
2022
2024
2026
Starting date of the conference